\documentclass{scrartcl}
\KOMAoptions{paper=letter}

\usepackage[fixcref]{group}

\newcommand{\NP}{\mathrm{NP}}
\newcommand{\coNP}{\mathrm{coNP}}
\newcommand{\Ptime}{\mathrm{P}}
\newcommand{\FP}{\mathrm{FP}}
\newcommand{\ThetaTwo}{\Theta_2^{\mathrm{p}}}
\newcommand{\prob}[1]{\textup{\textsc{#1}}}
\newcommand{\R}{\mathcal{R}}
\newcommand{\C}{\mathcal{C}}
\newcommand{\K}{\mathcal{K}}
\newcommand{\neglit}[1]{\overline{#1}}

\title{The Computational Complexity of\\ Weak Saddles in Zero-Sum Games}

\author{%
Felix Brandt\\
Technical University of Munich
\and
Markus Brill\\
University of Oxford
\and
Warut Suksompong\\
National University of Singapore
}

\begin{document}

\maketitle

\begin{abstract}
In the early 1950s, Lloyd Shapley proposed a family of set-valued solution concepts for zero-sum games called \emph{saddles}. While Shapley gave a polynomial-time algorithm to compute strict saddles, the computational complexity of weak saddles in zero-sum games has remained open. 
\citet{BBFH09a} gave a polynomial-time algorithm for computing weak saddles in symmetric zero-sum games \emph{without} off-diagonal ties. In this note, we show that finding a weak saddle is $\NP$-hard in symmetric zero-sum games \emph{with} off-diagonal ties, as well as in win-lose games. Remarkably, the reductions hinge on the ordinal minimax theorem, which establishes a structural property of weak saddles in zero-sum games that seemed to open the door to a polynomial-time algorithm.
\end{abstract}

\section{Introduction}

\Citet{Shap53a,Shap53b,Shap64a} showed that the existence of saddle points in zero-sum games can be guaranteed without randomization by moving to minimal \emph{sets} of actions. A \emph{generalized saddle point} (GSP) is a pair of action sets such that every action outside a player's set is dominated by some action inside it, given that the opponent plays an action from his set; a \emph{saddle} is an inclusion-minimal GSP. Depending on the underlying notion of dominance, one obtains \emph{strict} and \emph{weak} saddles. Every zero-sum game has a unique strict saddle \citep{Shap64a}, and strict saddles of arbitrary games can be computed in polynomial time \citep{BBFH09a,BrBr12a}. Weak saddles need not be unique, but they are unique in confrontation games (symmetric zero-sum games without off-diagonal ties) \citep{DuLe96a}, where they can be computed in polynomial time by repeatedly solving linear programs \citep{BBFH09a,BrBr12a}. All weak saddles of a zero-sum game are interchangeable and equivalent \citep{BBS14b}, an ordinal analog of the minimax theorem \citep{vNeu28a}. For general two-player games, finding a weak saddle is $\NP$-hard and recognizing one is $\coNP$-complete \citep{BBFH10c}. An overview of these and related computational results is given by \citet{BBFHH09a}.

The complexity of computing weak saddles in zero-sum games has been posed as an open problem \citep{BrBr12a,BBS14b}. The same question for the variant of weak saddles based on dominance with at least one strict inequality was raised earlier \citep{BBFHH09a,BBFH10c}. In view of the strong structural properties of weak saddles in zero-sum games, and the fact that mixed saddle points---Nash equilibria---of zero-sum games can be found by linear programming, one might have hoped for an efficient algorithm. Strikingly, we show that the problem is $\NP$-hard, i.e., no such algorithm exists unless $\Ptime=\NP$ (\Cref{thm:main}).

Our reduction (\Cref{sec:results}) merely uses \emph{win-lose games}, i.e., zero-sum games with only two distinct payoffs. Ironically, although our result is stronger than the hardness result for unrestricted two-player games by \citet{BBFH10c}, the proof is much simpler. The reason is that the ordinal minimax theorem itself \citep{BBS14b} makes the reduction work: by equivalence, every weak saddle of the constructed game has as many columns as a weak saddle that encodes a satisfying assignment, and this forces every weak saddle to encode one. A symmetrized version of the construction yields the same result for symmetric win-lose-tie games, which correspond to weak tournaments. Since weak saddles of confrontation games can be computed efficiently, off-diagonal ties are what separates tractable from intractable symmetric zero-sum games. In weak tournaments, the weak saddles whose row and column sets coincide correspond to minimal covering sets with respect to Richelson shading, a variant of McKelvey covering \citep{Dugg11a}; hence, finding such a set is $\NP$-hard, whereas the minimal covering set based on McKelvey covering can be computed in polynomial time \citep{BrFi08b}. The ordinal minimax theorem also yields an upper bound (see the beginning of \Cref{sec:results}): in zero-sum games, weak saddles are exactly the weak GSPs of minimum size, so a weak saddle can be found by binary search with logarithmically many queries to an $\NP$ oracle that returns witnesses.

We have obtained further hardness results, which we omit from this note. These include deciding whether a zero-sum game contains a nontrivial weak saddle, recognizing weak saddles, and deciding whether an action is contained in some weak saddle. The latter problem is $\ThetaTwo$-complete.
In terms of computational complexity, finding a weak saddle is equivalent to finding a maximum clique of a graph (under metric reductions), and computing the common size of all weak saddles is as hard as computing the clique number. 
Finding a weak saddle remains $\NP$-hard if dominance requires at least one strict inequality \citep[as in][]{BBFHH09a,BBFH10c}, but this requires a different mechanism for encoding assignments in the reduction proof.

\section{Preliminaries}\label{sec:prelim}

A (finite two-player) zero-sum game is given by a payoff matrix $A=(a_{r,c})_{r\in\R,c\in\C}$; the row player receives $a_{r,c}$ and the column player $-a_{r,c}$. For $C\subseteq\C$, row $r_1$ \emph{weakly dominates} row $r_2$ with respect to $C$, written $r_1\ge_C r_2$, if $a_{r_1,c}\ge a_{r_2,c}$ for all $c\in C$. For $R\subseteq\R$, column $c_1$ weakly dominates column $c_2$ with respect to $R$, written $c_1\le_R c_2$, if $a_{r,c_1}\le a_{r,c_2}$ for all $r\in R$. A set $R_1$ of rows dominates a set $R_2$ of rows with respect to $C$, written $R_1\ge_C R_2$, if every $r_2\in R_2$ is dominated by some $r_1\in R_1$; analogously for columns.

For nonempty $R\subseteq\R$ and $C\subseteq\C$, the product $R\times C$ is a \emph{weak GSP} if $R\ge_C \R\setminus R$ and $C\le_R \C\setminus C$ (\emph{external stability}). A \emph{weak saddle} is a weak GSP that contains no other weak GSP.\footnote{\Citet{BBFH10c} call these objects \emph{very weak} saddles, whereas weak saddles there are based on dominance requiring at least one strict inequality. Our terminology follows \citet{Shap64a} and \citet{BBS14b}.} Since $\R\times\C$ is a weak GSP, weak saddles always exist. We call $|R|+|C|$ the \emph{size} of $R\times C$.

A zero-sum game is a \emph{win-lose game} if it has only two distinct payoffs, which we take without loss of generality to be $0$ and $1$, and a \emph{win-lose-tie game} if its payoffs are in $\{-1,0,1\}$. 

A zero-sum game is \emph{symmetric} if $\R=\C=:\mathcal{A}$ and $a_{x,y}=-a_{y,x}$ for all $x,y\in\mathcal{A}$, i.e., the payoff matrix is skew-symmetric. In particular, $a_{x,x}=0$ for all $x$. A \emph{confrontation game} is a symmetric zero-sum game with $a_{x,y}\neq0$ for all $x\neq y$. Symmetric win-lose-tie games correspond to weak tournaments: action $x$ \emph{beats} action $y$, and $y$ \emph{loses against} $x$, if $a_{x,y}=1$, and $x$ and $y$ \emph{tie} if $a_{x,y}=0$.

\section{Results}\label{sec:results}

The problem we consider is \prob{FindWeakSaddle}: given a zero-sum game, find one of its weak saddles.

The ordinal minimax theorem of \citet{BBS14b} states that all weak saddles of a zero-sum game are interchangeable and equivalent. Interchangeability means that whenever $R\times C$ and $R'\times C'$ are weak saddles, so are $R\times C'$ and $R'\times C$. In particular, all weak saddles have the same number of rows and the same number of columns \citep[Theorem~2(ii)]{BBS14b}, and hence the same size $k^*$. Since every weak GSP contains a weak saddle, every weak GSP has size at least $k^*$, and a weak GSP of size $k^*$ contains, and therefore equals, a weak saddle. Hence, weak saddles are not only inclusion-minimal weak GSPs but exactly the weak GSPs of minimum size, which yields an upper bound on the complexity of finding them.

Deciding whether a zero-sum game has a weak GSP $R\times C$ of size $|R|+|C|\le k$ is in $\NP$, since a weak GSP serves as a certificate and external stability can be checked in polynomial time. Since $2\le k^*\le|\R|+|\C|$, binary search finds the smallest $k$ for which the answer is positive, namely $k^*$, with $O(\log(|\R|+|\C|))$ such queries. One further query for $k=k^*$, answered with a witness, returns a weak GSP of size $k^*$, which is a weak saddle. Hence, \prob{FindWeakSaddle} is in $\FP^{\NP}[\log,\mathit{wit}]$, the class of functions computable in polynomial time with logarithmically many queries to an $\NP$ oracle that, on a positive answer, also returns a witness \citep[see][]{Kraj95a}. In particular, if $\Ptime=\NP$, a weak saddle can be computed in polynomial time, because then the $\NP$ oracle, including its witnesses, can be simulated in polynomial time.

We reduce \prob{Sat} to \prob{FindWeakSaddle} for win-lose games and for symmetric win-lose-tie games. In both cases, the weak saddles of the constructed game encode the satisfying assignments of the given formula whenever it is satisfiable.

Let $\varphi=K_1\wedge\dots\wedge K_m$ be a formula in conjunctive normal form over the variables $x_1,\dots,x_n$, and let $L=\{x_1,\neglit{x}_1,\dots,x_n,\neglit{x}_n\}$ be the set of literals, with $n\ge2$. We assume that every clause $K_j$ is nonempty, since a formula with an empty clause is unsatisfiable. We identify an assignment with the set $T\subseteq L$ of the $n$ literals it makes true, so that $T$ is a \emph{satisfying assignment} if $T\cap K_j\neq\emptyset$ for all $j\in[m]$. For a literal $t\in L$, we let $\neglit{t}$ denote the negation of $t$.

\subsection{Win-Lose Games}

We start with a simple observation about win-lose games.

\begin{lemma}\label{lem:hit}
Let $A$ be a win-lose game in which every row contains a $0$, and let $R\times C$ be a weak GSP of $A$. Then for every row $r$, there is a column $c\in C$ with $a_{r,c}=0$.
\end{lemma}

\begin{proof}
Let $r$ be a row. By external stability (or since $r\ge_C r$ if $r\in R$), there is $s\in R$ with $s\ge_C r$. Let $c'$ be a column with $a_{s,c'}=0$. Again by external stability (or with $c=c'$ if $c'\in C$), there is $c\in C$ with $c\le_R c'$, and in particular $a_{s,c}\le a_{s,c'}=0$. Hence $0\le a_{r,c}\le a_{s,c}=0$, so $a_{r,c}=0$.
\end{proof}

\paragraph{The game $B_\varphi$.} The rows of $B_\varphi$ are the \emph{variable rows} $V=\{\nu_1,\dots,\nu_n\}$ and the \emph{clause rows} $\K=\{\kappa_1,\dots,\kappa_m\}$. The columns are the literals $\ell\in L$. The payoffs are
\[
a_{\nu_i,\ell}=\begin{cases}0 & \text{if } \ell\in\{x_i,\neglit{x}_i\},\\ 1 & \text{otherwise,}\end{cases}
\qquad\qquad
a_{\kappa_j,\ell}=\begin{cases}0 & \text{if } \ell\in K_j,\\ 1 & \text{otherwise.}\end{cases}
\]
Thus, $B_\varphi$ is a win-lose game in which variable row $\nu_i$ loses exactly against the two literals of $x_i$, and clause row $\kappa_j$ loses exactly against the literals of $K_j$. Since every clause is nonempty, every row contains a $0$, so \Cref{lem:hit} applies: the column set of every weak GSP contains a literal of every variable and a literal of every clause. An example is given in \Cref{fig:binexample}.

\begin{figure}[tb]
\centering
\begin{tabular}{@{}l*{6}{c}@{}}
\toprule
 & $x_1$ & $\neglit{x}_1$ & $x_2$ & $\neglit{x}_2$ & $x_3$ & $\neglit{x}_3$ \\
\cmidrule(lr){2-3}\cmidrule(lr){4-5}\cmidrule(l){6-7}
$\nu_1$ & \cellcolor{gray!30}$0$ & $0$ & $1$ & \cellcolor{gray!30}$1$ & \cellcolor{gray!30}$1$ & $1$ \\
$\nu_2$ & \cellcolor{gray!30}$1$ & $1$ & $0$ & \cellcolor{gray!30}$0$ & \cellcolor{gray!30}$1$ & $1$ \\
$\nu_3$ & \cellcolor{gray!30}$1$ & $1$ & $1$ & \cellcolor{gray!30}$1$ & \cellcolor{gray!30}$0$ & $0$ \\
\addlinespace
$\kappa_1$ & $0$ & $1$ & $0$ & $1$ & $1$ & $1$ \\
$\kappa_2$ & $1$ & $0$ & $1$ & $1$ & $0$ & $1$ \\
$\kappa_3$ & $1$ & $1$ & $1$ & $0$ & $1$ & $0$ \\
\bottomrule
\end{tabular}
\caption{The game $B_\varphi$ for $\varphi=(x_1\vee x_2)\wedge(\neglit{x}_1\vee x_3)\wedge(\neglit{x}_2\vee\neglit{x}_3)$ (payoffs of the row player). The game has exactly four weak saddles, namely $R\times T$ with $R\in\{V,\K\}$ and $T\in\{\{x_1,\neglit{x}_2,x_3\},\{\neglit{x}_1,x_2,\neglit{x}_3\}\}$, in accordance with interchangeability. The column sets are precisely the two satisfying assignments of $\varphi$ (\Cref{thm:bin}). The shaded cells form the weak saddle $V\times\{x_1,\neglit{x}_2,x_3\}$.}
\label{fig:binexample}
\end{figure}

We first show that every satisfying assignment, together with the variable rows, forms a weak saddle.

\begin{lemma}\label{lem:binsel}
For every satisfying assignment $T$, the pair $V\times T$ is a weak saddle of $B_\varphi$.
\end{lemma}

\begin{proof}
For $i\in[n]$, let $t_i$ denote the literal of $x_i$ in $T$. Restricted to $T$, row $\nu_i$ has a $0$ at $t_i$ and $1$s elsewhere. We first verify external stability. Let $j\in[m]$ and choose $t_i\in T\cap K_j$. Then $\kappa_j$ also has a $0$ at $t_i$, so $\nu_i\ge_T\kappa_j$. Moreover, the two literals of every variable yield identical columns on $V$, so $t_i\le_V\neglit{t_i}$ for every $i\in[n]$.

For minimality, let $R\times C\subseteq V\times T$ be a weak GSP. By \Cref{lem:hit} applied to the variable rows, $C$ contains a literal of every variable, hence $C=T$. If some $\nu_i\notin R$, then $\nu_k\ge_T\nu_i$ for some $\nu_k\in R$ with $k\neq i$, which is impossible because $\nu_k$ has a $0$ at $t_k$, where $\nu_i$ has a $1$. Hence $R\times C=V\times T$.
\end{proof}

Since all weak saddles of a zero-sum game have the same number of columns, \Cref{lem:binsel} also determines the column sets of all other weak saddles of $B_\varphi$.

\begin{theorem}\label{thm:bin}
If $\varphi$ is satisfiable, then the column sets of the weak saddles of $B_\varphi$ are exactly the satisfying assignments of $\varphi$.
\end{theorem}

\begin{proof}
Let $R\times C$ be a weak saddle. By \Cref{lem:hit}, $C$ contains a literal of every variable and of every clause. Hence $|C|\ge n$, with equality only if $C$ is a satisfying assignment. By \Cref{lem:binsel}, some weak saddle has $n$ columns, so by equivalence \citep[Theorem~2(ii)]{BBS14b}, every weak saddle has $n$ columns, and $C$ is a satisfying assignment. Conversely, every satisfying assignment is the column set of a weak saddle by \Cref{lem:binsel}.
\end{proof}

If $\varphi$ is unsatisfiable, the first two sentences of the proof still apply and show that every weak saddle of $B_\varphi$ has more than $n$ columns.

\subsection{Symmetric Win-Lose-Tie Games}

In a symmetric zero-sum game, $a_{r,c'}\le a_{r,c}$ is equivalent to $a_{c',r}\ge a_{c,r}$. Hence a column $c'$ dominates a column $c$ with respect to a set $X$ of rows if and only if row $c'$ dominates row $c$ with respect to $X$ as a set of columns. Consequently, $R\times C$ is a weak GSP if and only if $C\times R$ is. Together with interchangeability, this yields a simple description of the weak saddles of symmetric zero-sum games.

\begin{proposition}\label{prop:symform}
Let $A$ be a symmetric zero-sum game. Then $R\times C$ is a weak saddle of $A$ if and only if $R\times R$ and $C\times C$ are weak saddles of~$A$. Moreover, for $X\subseteq\mathcal{A}$, the pair $X\times X$ is a weak saddle if and only if $X$ is inclusion-minimal among all nonempty sets $Y$ such that every action outside $Y$ is dominated with respect to $Y$ by some action in $Y$.
\end{proposition}

\begin{proof}
If $R\times C$ is a weak saddle, then so is $C\times R$, since the map $R\times C\mapsto C\times R$ preserves weak GSPs and inclusion. By interchangeability \citep[Theorem~2(i)]{BBS14b}, $R\times R$ and $C\times C$ are weak saddles. The converse direction is again interchangeability. For the second statement, observe that the row and column conditions for $Y\times Y$ coincide, so $Y\times Y$ is a weak GSP if and only if $Y$ has the stated property. If $X\times X$ is a weak saddle, $X$ is clearly minimal with this property. Conversely, suppose that $X$ is minimal with this property. Then $X\times X$ is a weak GSP and contains some weak saddle $R\times C$. By the first statement, $R\times R\subseteq X\times X$ is a weak GSP, so $R=X$ by minimality, and likewise $C=X$.
\end{proof}

\paragraph{Covering sets.} For weak tournaments,\footnote{Recall that in a \emph{weak tournament}, for any two distinct actions $x$ and $y$, exactly one of the following holds: $x$ beats $y$, $y$ beats $x$, or $x$ and $y$ tie. A \emph{tournament} excludes the last possibility.} \Cref{prop:symform} relates weak saddles to covering sets from social choice theory. For a set $Z$ of actions and $x,y\in Z$, the condition $x\ge_Z y$ means that every action in $Z$ that $y$ beats is beaten by $x$ and every action in $Z$ that beats $x$ also beats $y$; in particular, $y$ does not beat $x$. \Citet{Dugg11a} calls this relation \emph{Richelson shading}; we say that $x$ \emph{shades} $y$ in $Z$. \emph{McKelvey covering} additionally requires that $x$ beats $y$. The minimal covering set of a weak tournament, which is defined in terms of McKelvey covering, is unique \citep{DuLa99a,PeSu99a} and can be computed in polynomial time \citep{BrFi08b}.

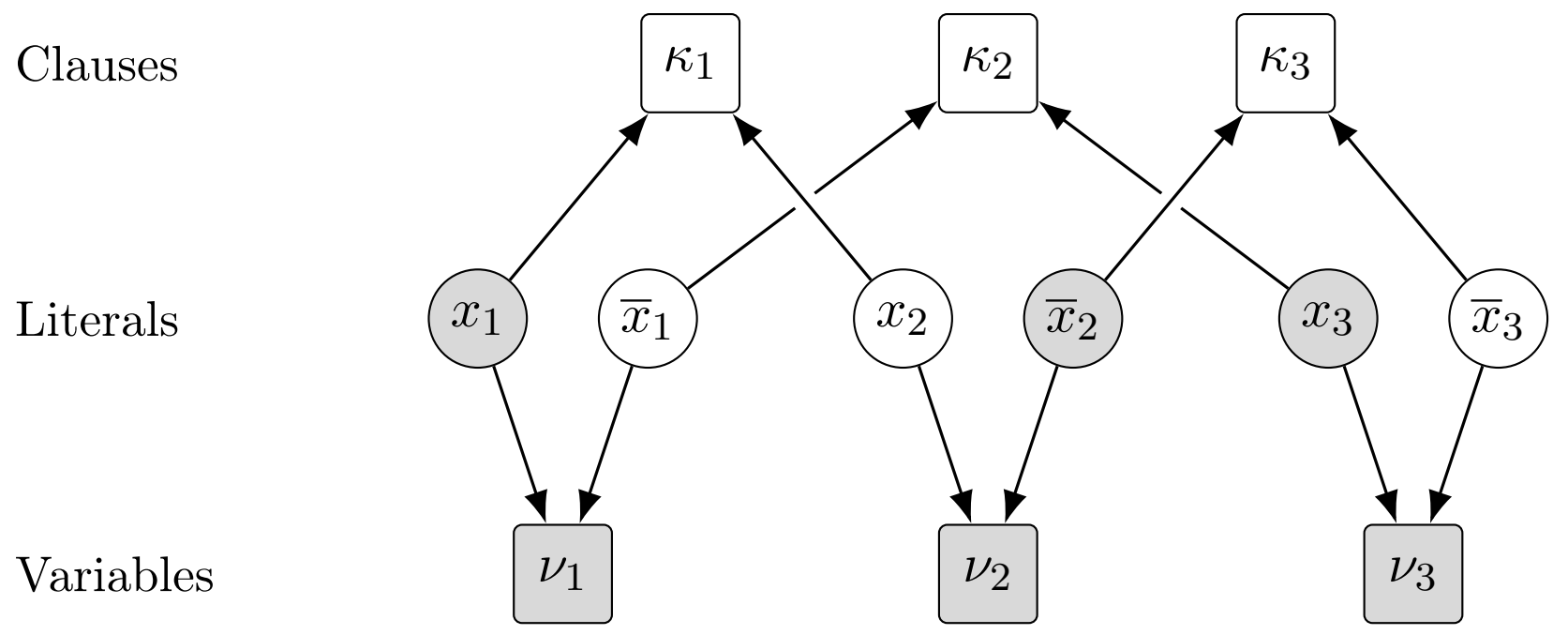
\begin{figure}[tb]
\centering
\begin{tikzpicture}[x=2.6em,y=2.6em,
  lit/.style={circle,draw,minimum size=1.8em,inner sep=0pt},
  box/.style={rectangle,rounded corners=0.15em,draw,minimum width=1.8em,minimum height=1.8em,inner sep=0pt},
  sad/.style={fill=gray!30},
  win/.style={-{Latex[length=0.6em]},semithick},
  bridge/.style={preaction={draw,white,line width=0.45em,-,shorten <=0.9em,shorten >=0.9em}},
  lab/.style={anchor=west,font=\small}]
\node[lab] at (-3.4,1.8) {Clauses};
\node[lab] at (-3.4,0) {Literals};
\node[lab] at (-3.4,-1.8) {Variables};
\node[lit,sad] (x1) at (0,0) {$x_1$};
\node[lit] (nx1) at (1.2,0) {$\neglit{x}_1$};
\node[lit] (x2) at (3,0) {$x_2$};
\node[lit,sad] (nx2) at (4.2,0) {$\neglit{x}_2$};
\node[lit,sad] (x3) at (6,0) {$x_3$};
\node[lit] (nx3) at (7.2,0) {$\neglit{x}_3$};
\node[box,sad] (v1) at (0.6,-1.8) {$\nu_1$};
\node[box,sad] (v2) at (3.6,-1.8) {$\nu_2$};
\node[box,sad] (v3) at (6.6,-1.8) {$\nu_3$};
\node[box] (k1) at (1.5,1.8) {$\kappa_1$};
\node[box] (k2) at (3.6,1.8) {$\kappa_2$};
\node[box] (k3) at (5.7,1.8) {$\kappa_3$};
\foreach \l/\v in {x1/v1,nx1/v1,x2/v2,nx2/v2,x3/v3,nx3/v3}
  \draw[win] (\l) -- (\v);
\draw[win] (x1) -- (k1);
\draw[win] (nx1) -- (k2);
\draw[win,bridge] (x2) -- (k1);
\draw[win] (x3) -- (k2);
\draw[win,bridge] (nx2) -- (k3);
\draw[win] (nx3) -- (k3);
\end{tikzpicture}
\caption{The game $S_\varphi$ for the formula $\varphi=(x_1\vee x_2)\wedge(\neglit{x}_1\vee x_3)\wedge(\neglit{x}_2\vee\neglit{x}_3)$ of \Cref{fig:binexample}, depicted as a weak tournament. Arcs are drawn from winners to losers. Each variable or clause action (box) beats every literal it is not joined to, any two boxes tie, and so do any two literals. The game has exactly $16$ weak saddles, namely $X\times Z$ with $X,Z\in\{V\cup T: T\in\mathcal{T}\}\cup\{\K\cup T: T\in\mathcal{T}\}$, where $\mathcal{T}=\{\{x_1,\neglit{x}_2,x_3\},\{\neglit{x}_1,x_2,\neglit{x}_3\}\}$, in accordance with interchangeability. The literal parts of all weak saddles are satisfying assignments of $\varphi$ (\Cref{thm:symbin}). The filled vertices form the set $V\cup T$ for $T=\{x_1,\neglit{x}_2,x_3\}$, and $(V\cup T)\times(V\cup T)$ is a weak saddle (\Cref{lem:symbinsel}). Every other action $y$ is shaded in $V\cup T\cup\{y\}$ by a filled vertex: $\neglit{x}_1$, $x_2$, and $\neglit{x}_3$ by $x_1$, $\neglit{x}_2$, and $x_3$, and $\kappa_1$, $\kappa_2$, and $\kappa_3$ by $\nu_1$, $\nu_3$, and $\nu_2$, respectively.}
\label{fig:symexample}
\end{figure}

Now let $X$ be a set of actions, $x\in X$, and $y\notin X$. Then $x\ge_X y$ is equivalent to $x\ge_{X\cup\{y\}}y$, since $a_{x,x}=0\ge a_{y,x}$ implies $a_{x,y}\ge0=a_{y,y}$. By \Cref{prop:symform}, $X\times X$ is thus a weak saddle if and only if $X$ is inclusion-minimal among the nonempty sets $Y$ that are \emph{externally stable}, i.e., every action $y\notin Y$ is shaded in $Y\cup\{y\}$ by an action in $Y$. Such a minimal set $X$ is also \emph{internally stable}, i.e., no action $x\in X$ is shaded in $X$ by another action $x'\in X$; otherwise, $X\setminus\{x\}$ would still be externally stable, since $\ge_X$ is transitive. Because every externally stable set contains a minimal one, the sets $X$ for which $X\times X$ is a weak saddle are exactly the minimal sets that are both internally and externally stable, i.e., the minimal covering sets \citep{Dutt88a,DuLa99a,PeSu99a} with McKelvey covering replaced by Richelson shading. In tournaments, the two relations coincide for distinct actions, which, together with the uniqueness of the minimal covering set of a tournament \citep{Dutt88a} and the first statement of \Cref{prop:symform}, yields the result of \citet{DuLe96a} that the weak saddle of a tournament is $X\times X$ for its minimal covering set $X$. By contrast, \Cref{thm:main} below and this characterization imply that finding a minimal covering set with respect to Richelson shading is $\NP$-hard. The characterization also makes it easy to verify weak saddles in drawings of weak tournaments such as \Cref{fig:symexample}.

\paragraph{The game $S_\varphi$.} The game $S_\varphi$, a symmetrization of $B_\varphi$, is the weak tournament on $V\cup\K\cup L$ in which the variable action $\nu_i$ loses against the two literals of $x_i$ and beats all other literals, the clause action $\kappa_j$ loses against the literals of $K_j$ and beats all other literals, any two actions in $V\cup\K$ tie, and so do any two literals. Hence, the payoffs of $V\cup\K$ against $L$ are those of $B_\varphi$ with $0$ replaced by $-1$. An example is given in \Cref{fig:symexample}.

\begin{lemma}\label{lem:symbinsel}
For every satisfying assignment $T$, the pair $(V\cup T)\times(V\cup T)$ is a weak saddle of $S_\varphi$.
\end{lemma}

\begin{proof}
For $i\in[n]$, let $t_i$ denote the literal of $x_i$ in $T$. By \Cref{prop:symform}, we have to show that $X=V\cup T$ is inclusion-minimal among the nonempty sets $Y$ such that every action outside $Y$ is dominated with respect to $Y$ by an action in $Y$. First, $X$ has this property: the literal $\neglit{t_i}$ yields the same payoffs as $t_i$ against $X$, and a clause action $\kappa_j$ is dominated by $\nu_k$ for any $t_k\in T\cap K_j$, since $\nu_k$ yields $-1$ against $X$ only at $t_k$, where $\kappa_j$ yields $-1$ as well.

Now let $Y\subseteq X$ be nonempty with this property, and let $I=\{k:\nu_k\in Y\}$ and $J=\{i:t_i\in Y\}$. If $i\in I\setminus J$, then $t_i$ yields $+1$ against $\nu_i\in Y$, whereas every action of $X$ other than $t_i$ yields at most $0$ against $\nu_i$; so $t_i$ is not dominated. Hence $I\subseteq J$. Next, let $k\notin I$; then $\nu_k$ must be dominated with respect to $Y$ by an action in $Y$. An action $\nu_l\in Y$ does not qualify: $l\in I\subseteq J$, and $\nu_l$ yields $-1$ against $t_l\in Y$, whereas $\nu_k$ yields $+1$. A literal $t_l\in Y$ yields $-1$ against $\nu_h$ for $h\neq l$ and $0$ against $T$, whereas $\nu_k$ yields $0$ against $V$ and $+1$ against $t_i$ for $i\neq k$; so $t_l$ qualifies only if $I\subseteq\{l\}$ and $J\subseteq\{k\}$. Since $t_l\in Y$, we have $l\in J\subseteq\{k\}$, so $l=k$ and $J=\{k\}$; together with $I\subseteq J$ and $k\notin I$, this gives $I=\emptyset$ and thus $Y=\{t_k\}$. Consequently, if $I\neq\emptyset$, then $I=[n]$ and thus $J=[n]$, i.e., $Y=X$. If $I=\emptyset$, then $Y=\{t_k\}$ for every $k\in[n]$, which is impossible since $n\ge2$.
\end{proof}

Using interchangeability, we can pair every weak saddle of $S_\varphi$ with the weak saddles of \Cref{lem:symbinsel}.

\begin{theorem}\label{thm:symbin}
If $\varphi$ is satisfiable, then for every weak saddle $X\times Y$ of $S_\varphi$, both $X\cap L$ and $Y\cap L$ are satisfying assignments of $\varphi$, and every satisfying assignment arises in this way.
\end{theorem}

\begin{proof}
Let $T$ be a satisfying assignment, and let $t_i$ denote the literal of $x_i$ in $T$ for $i\in[n]$. Since $X\times Y$ and $Y\times X$ are weak saddles, interchangeability and \Cref{lem:symbinsel} imply that $(V\cup T)\times Z$ and $Z\times(V\cup T)$ are weak saddles for $Z\in\{X,Y\}$, and by equivalence, $|Z|=|V\cup T|=2n$. Thus, every action $c\notin Z$ satisfies $z\ge_{V\cup T}c$ for some $z\in Z$.
\begin{itemize}
\item For a literal of $x_i$, an action $z$ with this property must yield $+1$ against $\nu_i$. Variable and clause actions yield $0$ against $V$, and literals of other variables yield $-1$ against $\nu_i$. Hence $Z$ contains $x_i$ or $\neglit{x}_i$.
\item For $\nu_k$, a literal does not qualify, since it yields $-1$ against all variable actions except the one of its own variable (recall $n\ge2$), whereas $\nu_k$ yields $0$ against $V$; neither does $\nu_l$ with $l\neq k$, which yields $-1$ against $t_l$. A clause action $\kappa_j$ yields $-1$ against the literals in $K_j$, whereas $\nu_k$ yields $+1$ against $t_i$ for $i\neq k$; so $\kappa_j$ qualifies only if $T\cap K_j\subseteq\{t_k\}$. Since $T$ is a satisfying assignment, $T\cap K_j\neq\emptyset$, so this means $T\cap K_j=\{t_k\}$. Hence $Z$ contains $\nu_k$ or a clause action $\kappa_j$ with $T\cap K_j=\{t_k\}$.
\end{itemize}
The $2n$ sets of actions identified above are pairwise disjoint. Since $|Z|=2n$, $Z$ contains exactly one action from each of them; in particular, $Z\cap L$ contains exactly one literal of every variable. Finally, every clause action $\kappa_j$ must be dominated with respect to $Z$ by an action in $V\cup T$. If $Z\cap L\cap K_j=\emptyset$, then $\kappa_j$ yields $+1$ against all of $Z\cap L$, whereas every variable action yields $-1$ against the literal of its variable in $Z\cap L$ and every literal yields $0$ against $Z\cap L$, a contradiction. Hence $Z\cap L$ satisfies $\varphi$. The last statement of the theorem follows from \Cref{lem:symbinsel}.
\end{proof}

Together with the upper bound at the beginning of \Cref{sec:results}, \Cref{thm:bin,thm:symbin} yield our main result.

\begin{theorem}\label{thm:main}
\prob{FindWeakSaddle} is $\NP$-hard under polynomial-time Turing reductions, even for win-lose games and for symmetric win-lose-tie games. Moreover, a weak saddle of a zero-sum game can be computed in polynomial time if and only if $\Ptime=\NP$.
\end{theorem}

\begin{proof}
Let $G_\varphi$ be $B_\varphi$ or $S_\varphi$; both games can be constructed from $\varphi$ in polynomial time. By \Cref{thm:bin,thm:symbin}, if $\varphi$ is satisfiable, the sets $Y\cap L$ for weak saddles $X\times Y$ of $G_\varphi$ are exactly the satisfying assignments of $\varphi$. (In symmetric games, $X\times Y$ is a weak saddle if and only if $Y\times X$ is, so it does not matter whether we consider row sets or column sets.)

Compute some weak saddle $X\times Y$ of $G_\varphi$. Then $\varphi$ is satisfiable if and only if $Y\cap L$ is a satisfying assignment, which can be checked in polynomial time. The last statement of the theorem follows from the first one and the upper bound at the beginning of \Cref{sec:results}.
\end{proof}

\section*{Acknowledgments}

This material is based on work supported by the Deutsche Forschungsgemeinschaft under grant BR 2312/14-1, by the Ministry of Education, Singapore, under the Academic Research Fund Tier 1 (FY 2026) grant number 251RES2604, and by an NUS Start-up Grant. The results of this note were found by ChatGPT Astra and Claude Opus under the authors' guidance. The original write-up was produced by Claude Opus and then improved by the authors.

\def\bibfont{\small}
\bibliography{group}

\end{document}